\documentclass{article}
\usepackage{spconf,amsmath,graphicx,hyperref}
\usepackage{cite}
\usepackage{amsmath,amssymb,amsfonts}
\usepackage{algorithmic}
\usepackage{graphicx}
\usepackage{textcomp}
\usepackage{xcolor}
\usepackage{array}
\usepackage{booktabs}
\usepackage{subfig}
\usepackage{stfloats}
\title{Q-A3C\textsuperscript{2}: Quantum Reinforcement Learning with Time-Series Dynamic Clustering for Adaptive ETF Stock Selection}
\name{Yen-Ku Liu$^{1}$, Yun-Cheng Tsai$^{1}$, Samuel Yen-Chi Chen$^{2}$}
\address{
$^{1}$Department of Technology Application and Human Resource Development,\\
National Taiwan Normal University, Taipei, Taiwan\\
$^{2}$Wells Fargo Bank, USA
}
\begin{document}
%
\maketitle
\begin{abstract}
Traditional ETF stock selection methods and reinforcement learning models such as the Asynchronous Advantage Actor–Critic (A3C) often suffer from high-dimensional feature spaces and overfitting when applied to complex financial markets. Moreover, static clustering algorithms fail to capture evolving market regimes, as the cluster with higher returns in one period may not remain optimal in the next. To address these limitations, this paper proposes \textbf{Q-A3C\textsuperscript{2}}, a quantum-enhanced A3C framework that integrates time-series dynamic clustering. By embedding Variational Quantum Circuits (VQCs) into the policy network, Q-A3C\textsuperscript{2} enhances nonlinear feature representation and enables adaptive decision-making at the cluster level. Experimental results on S\&P~500 constituents show that Q-A3C\textsuperscript{2} achieves a cumulative return of \textbf{17.09\%}, outperforming the benchmark’s \textbf{7.09\%}, demonstrating superior adaptability and exploration in dynamic financial environments.
\end{abstract}

\section{Introduction}
\label{sec:intro}

Exchange-traded funds (ETFs) are typically designed to passively replicate an index by rebalancing constituent stocks on a quarterly or semiannual basis. Yet, financial markets evolve rapidly—sector leadership, liquidity, and risk structures shift on much shorter timescales. Consequently, traditional ETF rebalancing often lags behind new industry trends and emerging risk regimes~\cite{liebi2020effect, wang2023etf, madhavan2016exchange}. 

Active fund managers, though more flexible, tend to hold concentrated portfolios. Empirical evidence shows such concentration can harm risk-adjusted returns during volatile periods, highlighting the need for diversification and adaptive allocation~\cite{chen2015concentration}. In response, increasing studies employ machine learning and reinforcement learning to dynamically capture market opportunities and adjust portfolios in a data-driven manner~\cite{bartram2021machine, Li_2021, Liu_2021}. 

However, most clustering-based portfolio strategies remain static, struggling to reflect evolving market regimes. As market microstructures and inter-asset relationships shift, static clustering fails to adapt, causing mismatches between model assumptions and real-world dynamics~\cite{CHOI2024112567}. This limitation motivates a \textit{time-rolling} or dynamic clustering approach, where clusters update continuously with market changes.

Recent advances in quantum machine learning (QML) further expand this direction. Variational quantum circuits (VQCs) embed classical data into high-dimensional Hilbert spaces via quantum feature maps, producing nonlinear embeddings analogous to kernel transformations~\cite{schuld2019quantum, doosti2024briefreviewquantummachine, cheng2024quantum}. Hybrid quantum–classical architectures have shown superior accuracy even on current noisy intermediate-scale quantum (NISQ) hardware~\cite{havlicek2019supervised}. Quantum reinforcement learning has also been explored for portfolio optimization and option pricing, showing performance gains despite hardware constraints~\cite{ORUS2019100028}.

Building on this foundation, Chen et al.\ proposed a quantum-enhanced asynchronous advantage actor–critic (A3C) model leveraging VQCs, achieving better results than classical baselines~\cite{chen2023asynchronous}. Likewise, Liu et al.\ found quantum A3C to exhibit strong sensitivity in volatile ETF trading, showing potential for dynamic decision-making~\cite{liu2025quantum}.

Motivated by these insights, this study examines whether quantum-enhanced reinforcement learning can further improve ETF stock selection. Traditional methods face two key challenges: (1) pure A3C suffers from high-dimensional feature inefficiency and overfitting; and (2) standalone clustering cannot ensure the best current cluster remains optimal next period. 

To address these, we propose \textbf{Q-A3C\textsuperscript{2}}—an integrated framework embedding clustering within A3C to enable rolling, adaptive decision-making. Leveraging quantum feature mapping for dimensionality compression and regime adaptation, Q-A3C\textsuperscript{2} effectively captures dynamic market structures, achieving improved trading performance and consistent outperformance over benchmarks such as the S\&P~500.

\section{Methodology}
\label{sec:methodology}

\begin{figure}[htbp]
\centering
\includegraphics[width=0.48\textwidth]{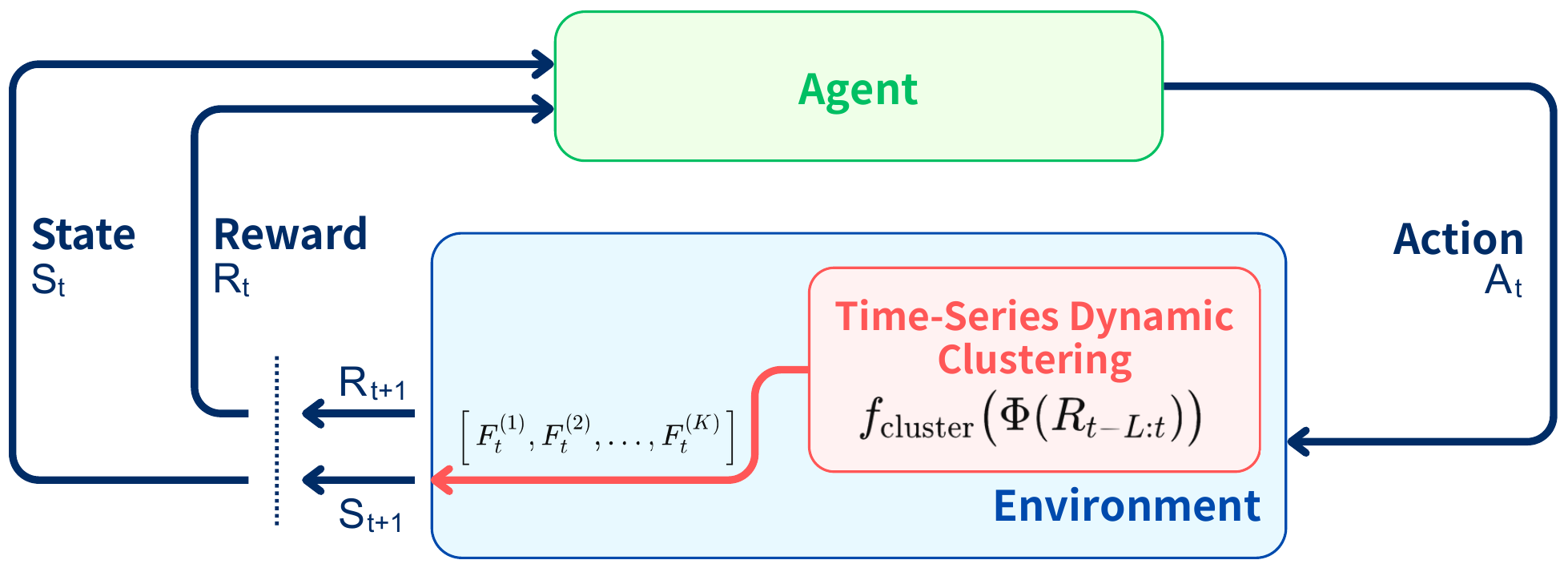}
\caption{Q-A3C\textsuperscript{2} framework with embedded Time-Series Dynamic Clustering module.}
\label{fig:q_a3c2_framework}
\end{figure}

We use the S\&P~500 as the benchmark for performance comparison~\cite{chavan2021intelligent}. Throughout the experiment, the proposed Q-A3C\textsuperscript{2} framework rebalances the investment portfolio on a rolling monthly basis, where all investment targets are drawn from the S\&P~500 constituent stocks. To capture the evolving market structure, we incorporate \textit{Time-Series Dynamic Clustering} into the reinforcement learning environment, allowing the agent to select clusters at the beginning of each month based on the market structure observed during the previous $L$-day window. The cluster returns in the current month are then used as feedback for learning~\cite{CHOI2024112567}. Figure~\ref{fig:q_a3c2_framework} illustrates the overall Q-A3C\textsuperscript{2} architecture, in which the dynamic clustering module is embedded within the A3C environment, enabling adaptive decision-making at the cluster level.

\textbf{Data and Feature Construction.}
We collected daily closing prices of all S\&P~500 constituent stocks from August~2021 to August~2024 for training, and from December~2024 to August~2025 for validation. To reflect market characteristics over multiple time horizons, we extract for each stock \(i\) at time \(t\) three representative features from the preceding $L$-day return window: the 5-day cumulative return \((m5_i(t))\), 20-day cumulative return \((m20_i(t))\), and 20-day return volatility \((v20_i(t))\)~\cite{Kong_2025}. These features jointly describe short-term momentum, mid-term momentum, and local volatility conditions—key drivers of cross-sectional asset behavior.

\textbf{Time-Series Dynamic Clustering and Market Structure.}
To reduce the high dimensionality of the stock space and adapt to regime shifts, a time-rolling clustering strategy is embedded into the environment. This design allows the agent to operate at the cluster level rather than the individual-stock level, reducing the feature dimensionality from the original \(N' \times 3\) to \(4 \times K + 2\).

At each decision window, K-Means clustering is applied to the feature matrix \(\Phi(R_{t-L:t}) \in \mathbb{R}^{N'\times3}\), where each row represents the three features \([m5_i(t), m20_i(t), v20_i(t)]\) for stock \(i\):
\begin{equation}
\label{eq:kmeans}
\boldsymbol{\ell}_t = f_{\mathrm{KMeans}}\!\left(\Phi(\mathbf{R}_{t-L:t})\right)
\in \{0,1,\ldots,K-1\}^{N'},
\end{equation}
where \(f_{\mathrm{KMeans}}(\cdot)\) denotes the K-Means operator. Each stock is assigned to one of \(K\) clusters, forming a cluster map that links tickers to cluster IDs. Each cluster \(C_t^{(k)}\) represents a distinct market behavior group whose collective return dynamics evolve over time~\cite{anugrahayu2023stock}.

\textbf{State Representation.}
We define the agent’s state at time \(t\) as:
\begin{equation}
\label{eq:state}
\mathbf{S}_t = \big[ \mathbf{F}_t^{(1)}, \ldots, \mathbf{F}_t^{(K)}, \mathbf{M}_t \big].
\end{equation}

\begin{equation}
\label{eq:clusterfeat}
\mathbf{F}_t^{(k)} =
\big[ m_{5,t}^{(k)},\; m_{20,t}^{(k)},\; v_{20,t}^{(k)},\; \mathrm{size}_{t}^{(k)} \big]^{\top},
\end{equation}
where $m_{5,t}^{(k)}$, $m_{20,t}^{(k)}$, and $v_{20,t}^{(k)}$ are the mean 5-day return, mean 20-day return, and mean 20-day volatility within cluster $C_t^{(k)}$, and $\mathrm{size}_t^{(k)}$ is the fraction of stocks assigned to the cluster.

\begin{equation}
\label{eq:marketfeat}
\mathbf{M}_t =
\big[ m_{5,t}^{(\mathrm{SPX})},\; m_{20,t}^{(\mathrm{SPX})} \big]^{\top},
\end{equation}

where \(m5_t^{(k)}\), \(m20_t^{(k)}\), and \(v20_t^{(k)}\) denote the mean 5-day return, 20-day return, and 20-day volatility across all stocks in cluster \(C_t^{(k)}\), and \(\text{size}_t^{(k)}\) represents the cluster’s relative size (ratio of constituent stocks). Thus, each cluster is compactly represented in four dimensions.  

\textbf{Action and Reward Design.}
At each decision step, the agent observes \(S_t\) and generates a cluster-selection policy via a \noindent\textbf{Quantum Feature Representation (VQC bottleneck).}
We insert a variational quantum circuit (VQC) as a nonlinear bottleneck in the policy network. Let
\begin{equation}
\label{eq:qbottleneck_in}
\mathbf{x}_t = \tanh\!\left(\mathbf{W}_2\,\tanh(\mathbf{W}_1\mathbf{S}_t + \mathbf{b}_1) + \mathbf{b}_2\right),
\end{equation}
where $\mathbf{S}_t$ is the state in \eqref{eq:state}. The VQC maps $\mathbf{x}_t$ to quantum features $\mathbf{u}_t=\mathrm{VQC}(\mathbf{x}_t;\boldsymbol{\theta})$, and the action logits are computed by
\begin{equation}
\label{eq:qbottleneck_out}
\mathbf{o}_t = \mathbf{W}_4\,\tanh\!\left(\mathbf{W}_3\,\mathbf{u}_t + \mathbf{b}_3\right) + \mathbf{b}_4.
\end{equation}
This explicit nesting addresses the network expression used in the implementation and clarifies the role of each layer.

For ablation, we implement a classical counterpart by replacing the VQC with a parameter-matched MLP bottleneck while keeping the same dynamic clustering environment and A3C training setup.
Action probabilities are obtained via a temperature-adjusted softmax with $\tau = 1.3$, promoting balanced exploration. Specifically,
\begin{equation}
\label{eq:policy}
\boldsymbol{\pi}_t = \mathrm{softmax}\!\left(\mathbf{o}_t/\tau\right), \qquad a_t \sim \boldsymbol{\pi}_t,
\end{equation}
where $\mathbf{o}_t$ are the logits in \eqref{eq:qbottleneck_out} and the sampled action $a_t \in \{1,2,\dots,K\}$ corresponds to investing in cluster $C_t^{(a_t)}$.
Initially, we explored a z-score-based reward, but it exhibited unstable learning under regime shifts and heavy-tailed returns (Fig.~\ref{fig:zscore_curve}). To stabilize training while keeping rewards comparable across months, we adopt a \textit{relative-optimality} reward:
\begin{equation}
\label{eq:reward}
r_t =
\begin{cases}
1, & \text{if } R_t^{(a_t)} = \max_j R_t^{(j)}, \\[4pt]
-\left(1 - \dfrac{R_t^{(a_t)} - \min_j R_t^{(j)}}{\max_j R_t^{(j)} - \min_j R_t^{(j)} + \epsilon}\right)^{2}, & \text{otherwise,}
\end{cases}
\end{equation}
where $R_t^{(j)}$ is the realized next-month return of cluster $j$ and $\epsilon$ is a small constant for numerical stability. This reward is bounded and scale-free: the best cluster receives $+1$, while suboptimal choices receive a smooth, quadratic penalty proportional to their relative gap. Empirically, this reduces collapse to a single mode and yields stable convergence (Fig.~\ref{fig:stability_curve}).  

Unlike conventional A3C settings that model the entire history as a single long episode, we treat each month as an episodic decision problem.
This design matches the practical ETF rebalancing cadence (monthly portfolio updates) and mitigates non-stationarity: market regimes and cross-asset relations can shift substantially across months, making a single continuous trajectory harder to learn and more sensitive to stale credit assignment.
Operationally, at the beginning of month $t$ the agent observes the market structure inferred from the most recent $L$ trading days and selects one cluster for allocation; the episode reward is then realized from the next-month cluster return.
Resetting the episode at each month therefore provides a controlled horizon, reduces long-range variance in returns, and aligns the learning signal with the intended deployment schedule.

\section{Experimental Results}
\label{sec:results}

This section presents the experimental outcomes of the proposed Q-A3C\textsuperscript{2} framework, including training behavior, trading performance, and cluster dynamics. All figures and tables are placed after first mention and formatted with compact float spacing.

\noindent\textbf{Baselines and Ablations.}
To isolate the effect of quantum feature representation from the embedded time-series dynamic clustering environment, we add a classical baseline that keeps the \emph{same} clustering-driven state/action/reward pipeline but replaces the VQC block with a parameter-matched classical module (MLP bottleneck) under the same A3C training setup.
In addition, to benchmark the contribution of \emph{dynamic} clustering, we compare against two traditional clustering strategies:
(i) \emph{Static clustering}, where K-Means is fitted once on the initial training window and reused without rolling updates;
(ii) \emph{Rolling clustering without RL}, where the agent greedily selects the cluster with the highest realized return in the previous month (a simple regime-following heuristic).
All methods share the same monthly rebalancing schedule and evaluation period for a controlled comparison.

\begin{figure}[!t]
\centering
\includegraphics[width=0.48\textwidth]{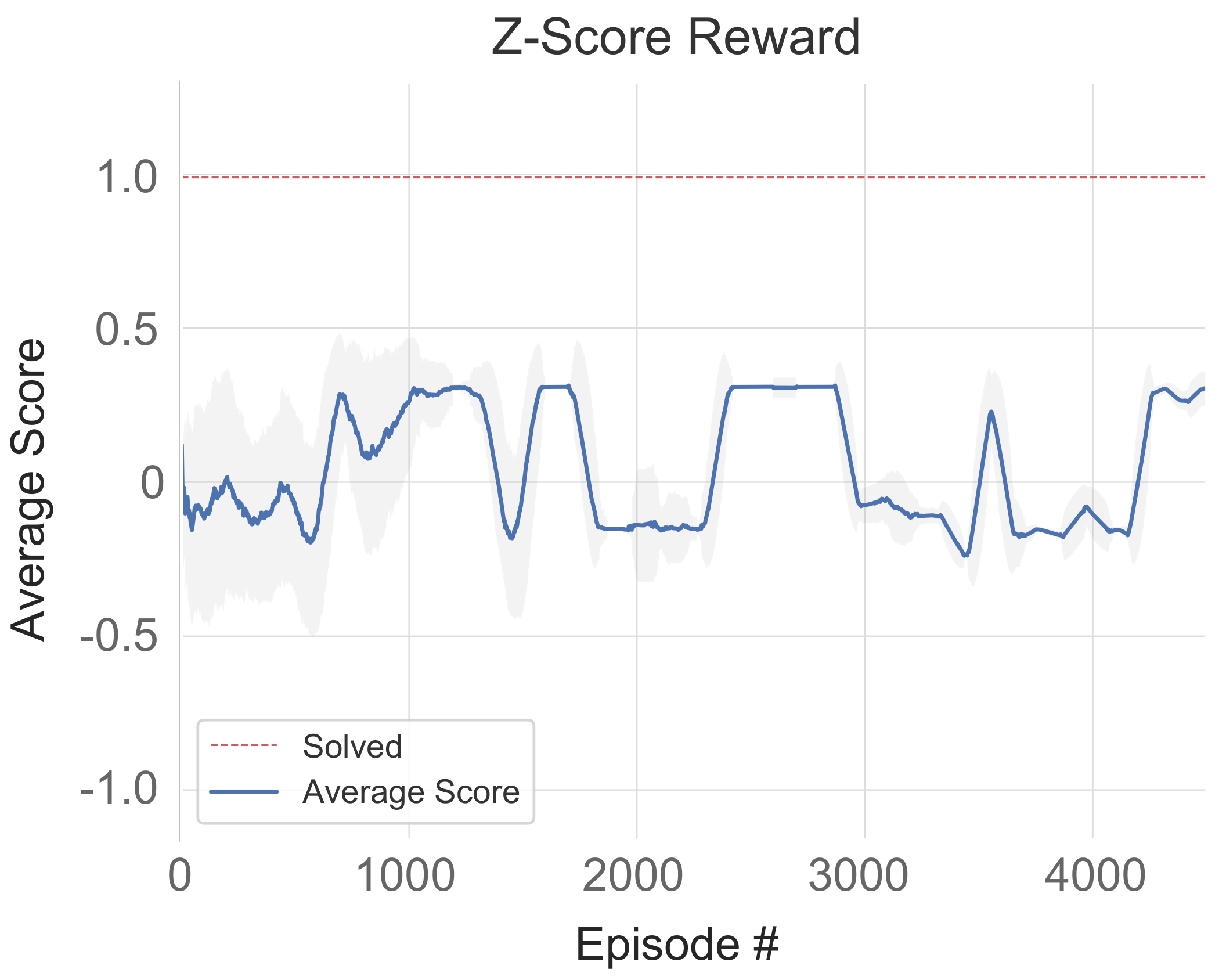}
\caption{Training performance under the Z-Score reward.}
\label{fig:zscore_curve}
\end{figure}

\begin{figure}[!t]
\centering
\includegraphics[width=0.48\textwidth]{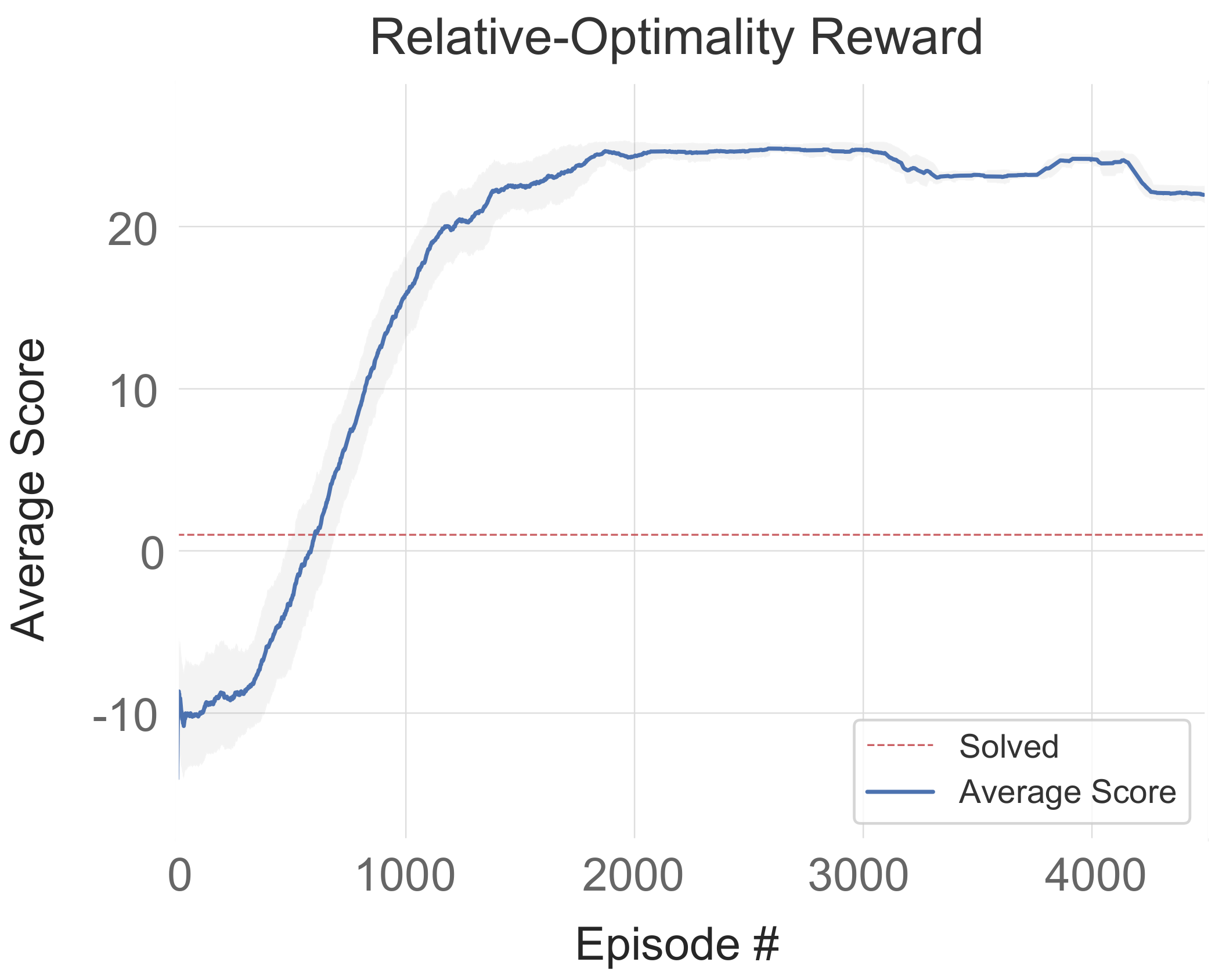}
\caption{Training performance under the Relative-Optimality reward.}
\label{fig:stability_curve}
\end{figure}

\textbf{Training Stability and Reward Mechanism.}  
Figure~\ref{fig:zscore_curve} and Figure~\ref{fig:stability_curve} compare the convergence behaviors of the model under the Z-Score and Relative-Optimality reward functions. When using the Z-Score reward, the agent tended to overfit to one or two clusters that frequently yielded moderate positive feedback, resulting in unstable and oscillatory performance. In contrast, the Relative-Optimality reward significantly improved convergence stability by penalizing non-optimal clusters according to their relative return distance. After approximately 2{,}000 epochs, the cumulative reward stabilized, indicating that the model had effectively learned a balanced exploration–exploitation strategy.

\begin{figure}[!t]
\centering
\includegraphics[width=0.48\textwidth]{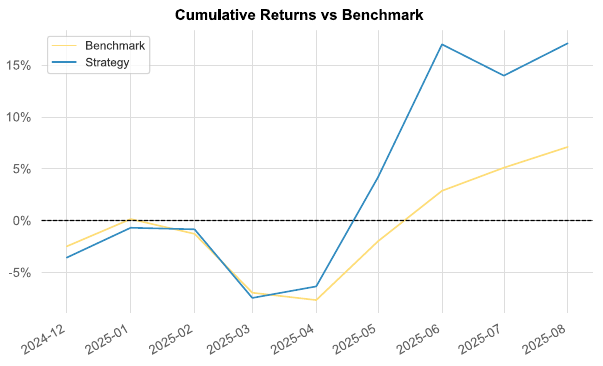}
\vspace{2mm}
\includegraphics[width=0.48\textwidth]{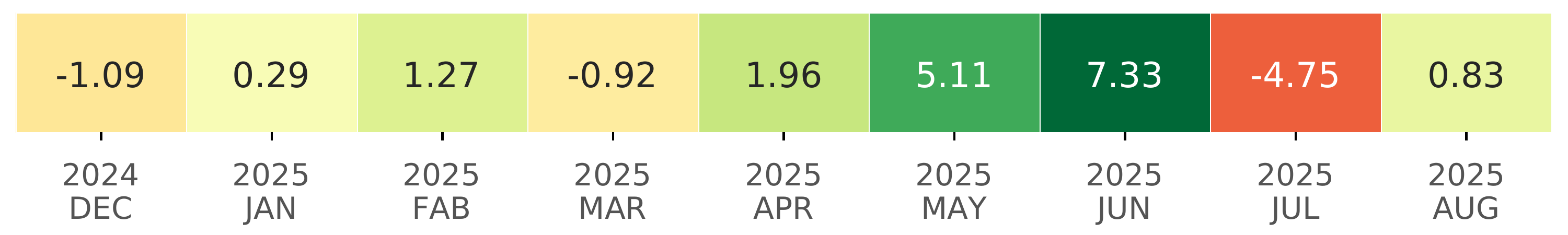}
\caption{Trading performance of Q-A3C\textsuperscript{2} vs.\ the S\&P~500 benchmark.}
\label{fig:trading_results_combined}
\end{figure}

\textbf{Trading Performance.}  
The model was trained from 2021-08-27 to 2024-08-31 for 4{,}500 epochs, and validated over the out-of-sample period from 2024-12 to 2025-08. The trading performance is benchmarked against the S\&P~500 index. As shown in Fig.~\ref{fig:trading_results_combined}, Q-A3C\textsuperscript{2} achieved a cumulative return of \textbf{17.09\%}, outperforming the benchmark’s \textbf{7.09\%}. The active return curve (strategy minus benchmark) illustrates consistent positive excess returns across multiple months, demonstrating the model’s ability to dynamically adjust its portfolio composition in response to changing market regimes. This result validates that integrating dynamic clustering within reinforcement learning can capture temporal shifts in market structure that traditional static strategies fail to exploit.

\textbf{Cluster Selection Dynamics.}  
To further understand the adaptive behavior of Q-A3C\textsuperscript{2}, Table~\ref{tab:stock_summary} summarizes the number of selected stocks and portfolio characteristics during the eight-month validation period. In the early stage (December~2024 to January~2025), the model concentrated on a small group of high-volatility growth stocks such as \texttt{TSLA} and \texttt{IBKR}, exhibiting a risk-seeking pattern. Beginning in February~2025, the number of selected stocks expanded to 135, reflecting a strong exploration phase in which the agent diversified across multiple sectors—information technology, healthcare, finance, and industrials—thereby enhancing exposure to cross-sector opportunities.

\begin{table*}[!t]
\centering
\small
\renewcommand{\arraystretch}{1.05}
\setlength{\tabcolsep}{3pt}
\caption{Monthly Summary of Selected Stocks and Portfolio Characteristics}
\begin{tabular}{@{} c c p{14.2cm} @{}}
\toprule
\textbf{Month} & \textbf{\# Stocks} & \textbf{Qualitative Summary} \\
\midrule
2024-12 & 10  & Focused on technology, energy, and financial stocks; portfolio small and tilted toward high-volatility growth targets. \\
2025-01 & 64  & Broader coverage across industries; entering a diffusion phase with wider exploration. \\
2025-02 & 135 & Peak diversity across nearly all sectors; highest level of industrial variety. \\
2025-03 & 60  & Re-concentration on information technology and industrials with clear growth characteristics. \\
2025-04 & 16  & Sharply reduced exposure; mainly travel, technology, and finance; refocus on high-volatility regions. \\
2025-05 & 1   & Extreme concentration on \texttt{PLTR} (AI-related stock); single high-confidence event-driven selection. \\
2025-06 & 20  & Re-diversification with technology (\texttt{NVDA}, \texttt{SMCI}) and energy sectors; rebound in risk appetite. \\
2025-07 & 107 & Expanded to industrials, utilities, and staples; rebalancing toward stable growth. \\
2025-08 & 96  & Sustained high diversity; gradual tilt toward defensive large-cap holdings. \\
\bottomrule
\end{tabular}
\label{tab:stock_summary}
\end{table*}

From a longitudinal perspective, certain stocks such as \texttt{LULU}, \texttt{IBKR}, \texttt{CCL}, and \texttt{UAL} reappeared across multiple months, suggesting that the model developed persistent preferences toward specific sectors—consumer discretionary, financials, and travel—aligned with the evolving macroeconomic cycle. By May~2025, the portfolio became extremely concentrated, selecting only \texttt{PLTR}, likely influenced by market volatility and AI-related momentum. From June onward, the model diversified again and increased exposure to defensive sectors such as utilities and consumer staples, indicating a shift toward risk control amid possible market corrections.

\begin{figure}[!t]
\centering
\includegraphics[width=0.48\textwidth]{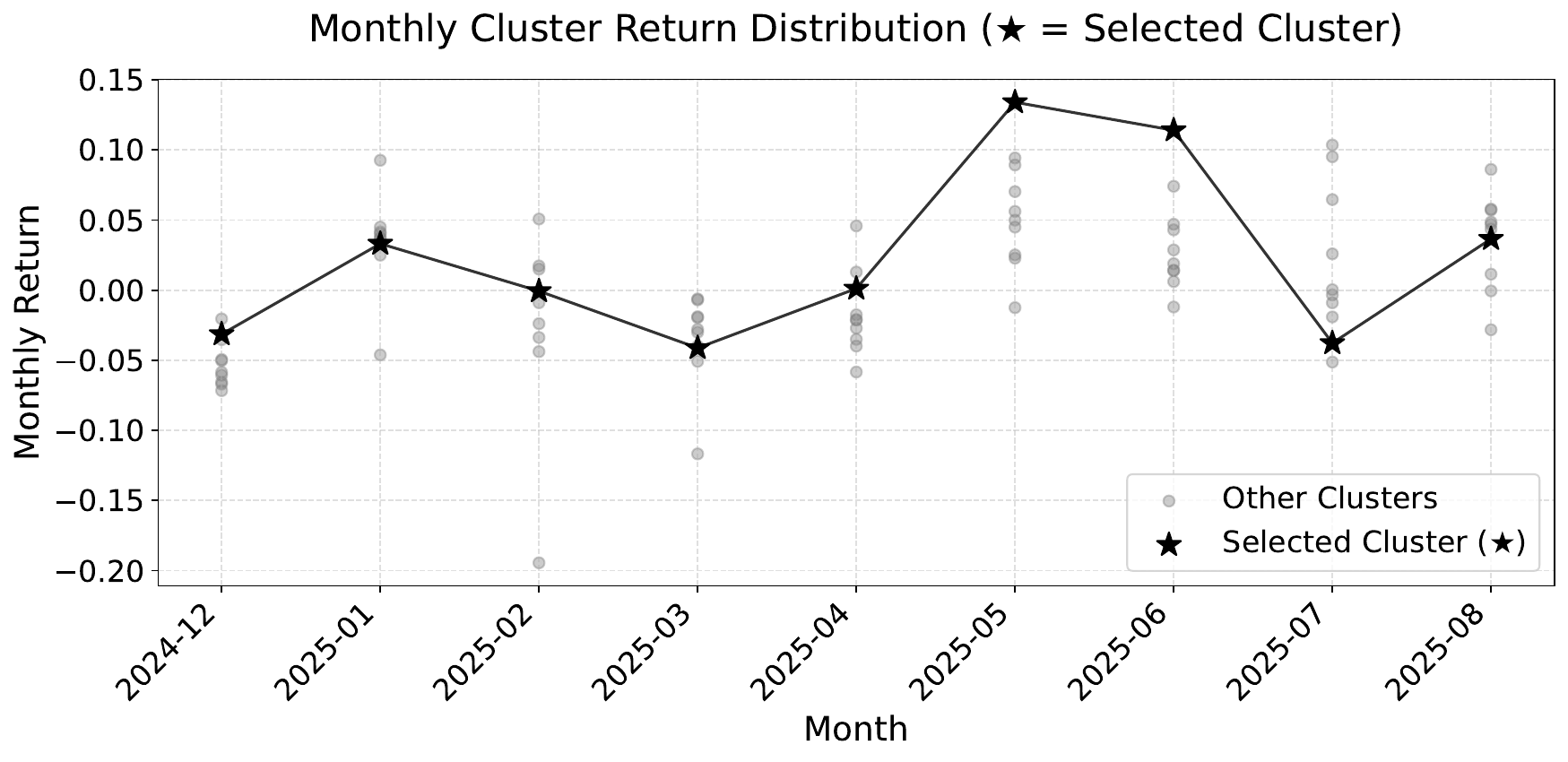}
\caption{Monthly cluster return distributions and selected cluster performance.}
\label{fig:cluster_return_dist}
\end{figure}

As shown in Fig.~\ref{fig:cluster_return_dist}, the agent consistently selected clusters that achieved higher subsequent-month returns, particularly in May and June~2025, where chosen clusters clearly outperformed others. These results demonstrate the effectiveness of the time-series dynamic clustering mechanism, enabling the model to recognize regime transitions and adjust its investment focus accordingly. Overall, Q-A3C\textsuperscript{2} illustrates the benefit of combining quantum-enhanced reinforcement learning with dynamic clustering—delivering higher cumulative returns, better adaptability, and stronger robustness under changing market conditions.

\section{Discussion and Conclusion}
Q-A3C\textsuperscript{2} addresses high-dimensional features and static clustering in ETF stock selection by embedding time-series dynamic clustering into the A3C loop~\cite{QIU2024122243}, enabling simultaneous feature compression and regime-adaptive decision-making. The proposed relative-optimality reward stabilizes training and mitigates over-concentration on short-term dominant clusters. With Variational Quantum Circuits, Q-A3C\textsuperscript{2} enhances nonlinear representation and shows improved responsiveness to market transitions, achieving 17.09\% cumulative return versus 7.09\% for the S\&P~500.Future work includes adaptive clustering~\cite{doshi2017towards}. Overall, Q-A3C\textsuperscript{2} enables adaptive portfolio optimization via quantum-enhanced RL.

\section*{Acknowledgment}
The views expressed in this article are those of the authors and do not represent the views of Wells Fargo. This article is for informational purposes only. Nothing contained in this article should be construed as investment advice. Wells Fargo makes no express or implied warranties and expressly disclaims all legal, tax, and accounting implications related to this article.

\bibliographystyle{IEEEbib}
\bibliography{references}

\end{document}